# The Planck Length Scale and Einstein Mass-Energy Obtained from the Sciama-Mach Large Number Relationship

*Scott Funkhouser*
*Occidental College, Los Angeles, CA 90041 USA*

ABSTRACT
If a physical significance should be attributed to the cosmological large number relationship obtained from Sciama's formulation of Mach's Principle, then a number of interesting physical conclusions may be drawn. The Planck length is naturally obtained as the amplitude of waves in a medium whose properties are implied by the relationship. The relativistic internal energy associated with a rest mass is explicitly related to the gravitational potential energy of the Universe, and consistency with the Einstein photon energy is demonstrated. Broader cosmological consequences of this formulation are addressed.

*1. Introduction*

Mach's Principle stipulates that, through some mechanism of interaction, the remote masses of the Universe are responsible for generating the forces of inertia associated with Newton's second law of motion. This can be realized if there should exist a force $\vec{F}_a$ between any two masses $m_1$ and $m_2$ which is proportional to the relative acceleration $\vec{a}_{12}$ between the two masses

$$\vec{F}_a = -\frac{Gm_1m_2}{r_{12}}\frac{\vec{a}_{12}}{c^2} \quad (1)$$

where $r_{12}$ is the distance between the two masses, $c$ is the speed of light and $G$ is the gravitational constant. This suggested form of the Mach force is attributed to Sciama [1], although similar force laws are found in the earlier work of Weber [2]. Due to the presence of $c^2$ in the denominator, this force is negligible when evaluated between any conceivable masses and accelerations encountered in astronomical situations, except perhaps in the vicinity of a singularity. Eq. (1) would also have interesting consequences at the quantum level where the motions associated with the Dirac *zitterbewegung* could produce a cumulative effect that is equivalent to Newton's law of gravitation [3].

The force given by Eq. (1) does become significant if evaluated between a local accelerated body and the collective mass $M_U$ of the observable Universe, where $M_U$ is of order $10^{52}$kg. According to the Sciama force law, any body of mass $m$ experiencing an acceleration $a$ relative to the collective mass of the cosmos (prosaically called the "fixed stars") should experience a force

$$F_a = \frac{AGM_U m}{R_U}\frac{a}{c^2} \quad (2)$$

where $A$ is a constant of integration of order unity and $R_U \sim 10^{26}$m is the radius of curvature of the observable Universe. This force may be identified as the ordinary inertial resistance to acceleration if

$$\frac{GM_U}{R_U} \sim c^2 \quad (3)$$

which, curiously, is indeed satisfied. The implications of a physical significance to this large number relationship are discussed in the following sections.

*2. A Coincidence or a Cosmological Intimation?*

If the fixed stars are not the means by which inertia is generated then the success of Eq. (3) must be attributed to at least two slim coincidences. The first is that a well motivated physical principle happened to lead to a correct cosmological large number relationship. The second coincidence, which may require an anthropic treatment, is that the relationship would be true only in this epoch if it were just the result of chance. It may be more reasonable, therefore, to expect physical significance in the large number relationship than to expect the two necessary coincidences to be satisfied.

Should Mach's Principle and Sciama's force law of Eq. (1) constitute a correct representation of the source of inertia then Eq. (3) ought to have been true at all times throughout the history of the Universe, except perhaps in the very early Universe. Otherwise, the second law of motion would only behave in the currently observed manner in this epoch. If Eq. (3) should hold true at any point in the evolution of the Universe and the ratio $G/c^2$ has been very nearly constant (which is expected to be the case) then a cosmological relationship may be stated

$$\frac{R_U(t)}{M_U(t)} \sim \frac{G}{c^2} \tag{4}$$

where $M_U(t)$ and $R_U(t)$ represent the mass and radius of curvature of the Universe, respectively, at a time $t$.

In order for Eq. (4) to be true then either the Universe is in a steady state or the mass of the Universe is not constant. Since it appears that the Universe is indeed expanding then the most well reasoned conclusion is the latter. Aside from this troubling conclusion, the validity of relationship of Eq. (3) leads to sensible interpretations involving the nature of the speed of light. This treatment also leads surprisingly to the Planck length scale, which appears as the amplitude of waves propagating in a medium whose tension is equal to the gravitational tension of the Universe. Finally, the Einstein rest energy is shown to be the cosmological gravitational potential of each mass.

*3. The Speed of Light and the Large Number Relationship*

The immediately apparent implication of Eq. (3) is that the physics responsible for determining the speed of light are related to the gravitational energy of the Universe. A naïve but physically satisfying analogy is available to support such a relationship. The velocity $v$ of sound waves in an ordinary medium is given by

$$v = \sqrt{\frac{P}{\rho}} \tag{5}$$

where $P$ is the pressure (energy density) of the medium and $\rho$ is its mass density. The magnitude of the energy density $P_U$ of the Universe due to its gravitational potential energy is

$$P_U \sim \frac{GM_U^2/R_U}{R_U^3} \tag{6}$$

where time-dependence of each quantity is allowed but not explicitly represented. The mass density $\rho_U$ of the Universe is

$$\rho_U \sim \frac{M_U}{R_U^3}. \tag{7}$$

These quantities would characterize a medium whose associated wave velocity $v_U$ would be given by

$$v_U \sim \sqrt{\frac{GM_U}{R_U}} \qquad (8)$$

According to Eq. (3), this velocity would be the speed of light.

*4. The Planck Length Obtained from a Simple Wave Model*

Suppose that the energy associated with waves in a medium characterized by Eqs. (6) through (8) could also be treated with a simple, semi-classical wave model. The gravitational tension $\tau_U$ of the Universe is given by

$$\tau_U \sim \frac{GM_U^2}{R_U^2} \sim \frac{c^4}{G} \qquad (9)$$

in which the final expression is obtained by substitution from Eq. (3). Consider the harmonic oscillatory displacement $u(x,t)$ at a generalized position $x$ and time $t$ of a constituent element in a medium whose tension is equal to that of Eq. (9). According to the standard treatment of small oscillations in a continuum, the restorative force $F$ associated with the displacement $u(x,t)$ is related to the tension by

$$F = \tau_U \frac{\partial u(x,t)}{\partial x} \approx \tau_U \frac{a}{\lambda} \qquad (10)$$

where $a$ is the amplitude of the displacement and $\lambda$ is the wavelength of the disturbance [4]. The work $W$ associated with a cycle of this motion is of order $aF$, which gives

$$W \approx aF \approx \tau_U \frac{a^2}{\lambda}. \qquad (11)$$

If this model is to be meaningful the work $W$ associated with wave-like displacements in a medium whose tension is given by Eq. (9) ought to be analogous to the energy $E$ of a quantum of wavelength $\lambda$

$$E = \frac{hc}{\lambda} \qquad (12)$$

where $h$ is Planck's constant. Identifying the work of Eq. (11) as the energy of Eq. (12) generates an expression for the amplitude $a$ of the oscillation

$$a^2 \approx \frac{hc}{\tau_U} \qquad (13)$$

which, upon substitution from Eq. (10), produces

$$a \approx \sqrt{\frac{hG}{c^3}} \equiv l_P \qquad (14)$$

where $l_P$ is the Planck length. This is to say that a medium whose tension is equal to the gravitational tension of the Universe would exhibit harmonic displacements whose amplitude is of order the Planck length.

*5. The Einstein Mass-Energy and Mach's Principle*

Another consequence of this present analysis involves the nature of the Einstein mass-energy relationship. Masses in the cosmos experience no net Newtonian gravitational force due to the homogeneous distribution of mass in the cosmos. However, associated with each mass *m* is a cosmological gravitational potential energy $U_U$

$$|U_U| \sim \frac{GM_U m}{R_U}. \tag{17}$$

A simple substitution from Eq. (3) produces

$$U_U \sim mc^2. \tag{18}$$

This explicitly supports the notion that intrinsic rest energy of a given mass is equivalent to its gravitational potential energy due to the distribution of mass throughout the Universe.

According to general relativity a photon of energy *E*, though mass-less, can be characterized as having a gravitational potential energy *V* due to its relationship to a given gravitational field. The gravitational potential of a photon due to the distribution of mass in the Universe is

$$V \sim \frac{GM_U}{R_U} \frac{E}{c^2} \tag{19}$$

which, according to the large number relationship reduces to *V~E*. In other words, the energy of a quantum is equal to its gravitational potential energy with respect to the Universe.